\documentclass{PoS}

\usepackage[numbers,sort&compress]{natbib}
\usepackage{multicol}
\usepackage[utf8]{inputenc}
\usepackage[T1]{fontenc}
\usepackage[english]{babel}
\usepackage{xspace}
\usepackage{amsmath}
\usepackage{wasysym}
\usepackage{lineno}

\newcommand\Fermi{\textsl{Fermi}\xspace}

\title{The \Fermi Sky in a Multimessenger Context}
\ShortTitle{The \Fermi Sky in a Multimessenger Context}

\author{\speaker{Felicia Krau{\ss}\,for the \Fermi-LAT collaboration}\\
GRAPPA \& Anton Pannekoek Institute for Astronomy, University of Amsterdam, Science Park 904, 1098 XH Amsterdam, The Netherlands\\}

\abstract{The \textsl{Fermi Gamma-Ray Space Telescope} has been
  launched on June 11 2008. While it has fundamentally changed our
  understanding of the high-energy $\gamma$-ray sky, it is even more
  powerful in multiwavelength and multimessenger efforts.
  In this work I summarize results from \Fermi-LAT in the
  multimessenger context, pertaining to high-energy neutrinos.
}
\FullConference{Neutrino Oscillation Workshop\\
  4 - 11 September, 2016\\
  Otranto (Lecce, Italy)}

\begin{document}

\section{Cosmic rays}

Earth undergoes a constant influx of charged particles and nuclei,
mostly protons. These cosmic rays have an $\sim 
E^{-2}$ spectrum with two 'breaks', at $\sim 10^7$ and $\sim
10^9$\,GeV, called the knee and the ankle,
respectively. Cosmic rays at energies up to the ankle are
typically thought to originate in Galactic sources, including (but not
limited to) supernovae, supernovae remnants and pulsar wind nebulae.
Supernovae are currently the only confirmed extrasolar neutrino
emitters \citep[SN\,1987A; for a review see][]{Arnett1989} and therefore contribute to the low energy
cosmic-ray spectrum \citep[for a review see][]{Blasi2013, Amato2014}.
As the cosmic-ray spectrum extends up to energies of $\sim
10^{21}$\,eV, a long standing questions is: which sources can accelerate
cosmic rays to these ultra-high energies?
Direct identification of the origin of cosmic-ray protons and heavier nuclei
is naturally hindered by magnetic fields: the Galactic
magnetic field, Earth's magnetic field, and possibly intergalactic
magnetic fields \citep[IGMF;][]{Biermann1997}. These deflect the
charged particles.


\subsection{Photopion production}
If an astrophysical source is capable of accelerating protons to very
high and ultra high energies a photon seed field is required at $\sim
10^{19}$\,eV, below which proton-proton interactions dominate. At
ultra-high energies the protons disintegrate in interactions with UV
to X-ray photons resulting in charged and neutral pions.
Neutral pions decay into a high-energy photon pair; this process is
called photopion production. These photons can have energies between
keV (X-ray) and TeV ($\gamma$-rays), with an energy of $\sim$ 10\% of
the proton energy, providing a hadronic explanation for
the high-energy emission that is seen in many Galactic and
extragalactic astrophysical objects
\citep{Mannheim1993,Mannheim1995,Mannheim1995b}. TeV $\gamma$ rays,
however, likely do not escape the dense environments and produce an
electron-positron pair. 
This makes high-energy photons in the keV to GeV energy range a
powerful tool for identifying sources of cosmic rays.
A detection of high-energy electromagnetic emission is not
evidence for hadronic processes, however, as the high-energy emission of many
sources can also be explained by the Inverse Compton process.
Low energy photons (e.g., produced by Synchrotron emission of 
relativistic electrons or from an external photon field) are
up-scattered by relativistic electrons.  
In many sources current leptonic and hadronic models can explain the
available data equally well \citep[e.g.,][]{Boettcher2013}. In some sources a certain
model is favored over another, but in general a broadband spectrum of
a source does not guarantee to be able to distinguish between leptonic
and hadronic scenarios.

\subsection{Expected neutrino emission}
In the hadronic case the charged pions decay into muons and neutrinos. The
muons then further produce neutrinos.
Assuming isospin symmetry, a correlation between the high-energy
electromagnetic flux (assuming it stems from the decay of neutral
pions) and high-energy neutrinos is expected with $F_\gamma=F_\nu$. An
association of high-energy neutrinos with a high energy source would
be unambiguous proof of the hadronic origin and contribute
significantly to the knowledge of cosmic ray emitters.
An association at a high statistical significance remains elusive but
is within reach in the next few years with increasing detections of
high-energy neutrino events.
%
%
In order to understand the neutrino signal, it is crucial to have
time-resolved information of the high-energy electromagnetic emission.

\section{\Fermi and the Large Area Telescope}

The \Fermi Gamma-ray Space Telescope has two instruments on board, the
Gamma-ray Burst Monitor (GBM) and the Large Area Telescope(LAT)
\citep{Atwood2009}. It is an ideal instrument to study possible
neutrino counterparts as it is continuously monitoring the sky,
providing spectral and flux information for the full sky since its
launch in 2008.  
The GBM monitors the sky at MeV energies in order to detect and study
Gamma-ray Bursts (GRBs) and other transient events such as Terrestrial
Gamma-ray Flashes \citep[TGFs;][]{Meegan2009}. 
The Large Area Telescope (LAT) operates at higher energies above $\sim
20$\,MeV. It is a pair conversion telescope with a combination of
silicon microstrip detectors and tungsten conversion foils at the top
of the instruments and a CsI scintillator calorimeter below.
It observes the entire sky in three hours \citep{Atwood2009}. Its
large effective area, angular resolution, energy range, and field of
view make it an ideal instrument to also study faint extragalactic
sources and search for counterparts of high-energy neutrinos.
The largest fraction of sources it detects are blazars, a subclass of
Active Galactic Nuclei (AGN) with a small angle between the line of
sight and their relativistic outflows, called jets.
Surprisingly, the second largest fraction of LAT sources \citep[in the
latest 3FGL catalog;][]{FGL3} are unassociated. It is generally believed
that these are largely blazars without known counterparts at other
wavelengths, though it is unclear what makes them $\gamma$-ray bright, 
but faint at other energies.

\section{IceCube detection of astrophysical neutrinos}

Much progress has been made in the last few years concerning the
detection of neutrinos, with the finalization of the IceCube detector
at the South Pole in 2010, and the ANTARES experiment in the
Mediterranean in 2008.
IceCube has detected $\sim$50 fully contained neutrino events, with a
significant excess above the expected atmospheric background. The spectrum
of atmospheric neutrinos is expected to fall steeply with increasing
energy, with only very few events expected above 100\,TeV and none above 1\,PeV.
I therefore only consider those PeV neutrinos as the best possible way to
identify astrophysical sources of cosmic rays
\citep{IceCube2013a,IceCube2013b,IceCube3year,IceCube4yr}.
Unfortunately the detection of high-energy neutrino events does not
lead to a direct association with a \Fermi-LAT counterpart, as these
neutrino events often have large angular uncertainties
($\sim10$-$15^\circ$ for shower-like $\nu_e$ events, and $\sim\,1^\circ$ for
$\nu_\mu$ track events). Further detections of muon track events with small
angular uncertainties offer a better possibility of a direct
association.
Additionally it has to be taken into account that a large number of
sources remains unresolved by \Fermi-LAT. We therefore do not expect an
association for each event. Studying the \Fermi-LAT extragalactic
background makes it possible to estimate the fraction of known
and unknown counterparts. We previously found that only 30\% of
neutrino events are expected to be associated with a LAT counterpart,
assuming blazars as the sources of the neutrinos \citep{fermisymp,bb}.

\section{Astrophysical sources of the neutrino signal}

In this section I briefly summarize candidates for the observed
neutrino emission and give the results that have come out of studies
with \Fermi-LAT data.

\subsection{Galactic sources}
For most Galactic high-energy emitters neutrino production has been
predicted. These include pulsars
\citep{Eichler1978,Harding1990,Anchordoqui2003,Link2005,Link2006,Bhadra2008},
X-ray and $\gamma$-ray binaries
\citep{Gaisser1985,Eichler1993,Levinson2001,Bednarek2005,Razzaque2010,Sitarek2012,Bednarek2014},
supernovae and pulsar wind nebulae
\citep{Cheng1990,Aharonian1996,Protheroe1998,Bednarek2003,Mandelartz2015,DiPalma2016},
and the Fermi bubbles \citep{Crocker2012,Lunardini2012,Razzaque2013,Lunardini2015}.
The LAT found evidence for pion decay signatures in the spectrum of
many supernova remnants
\citep{Abdo2009,Tanaka2011,Ajello2012,Giordano2012,Ackermann2013}.
While these likely largely contribute to the neutrino flux at lower
energies it is theoretically difficult to explain neutrinos up to PeV
energies, with the highest predicted energy at 300\,TeV.
It has further been argued that high-energy cosmic rays must originate from
outside of the Galactic disk as their Larmor motion can not be
contained \citep{Cocconi1956,Cocconi1996}.

\subsection{Extragalactic sources}
The observed distribution of all contained neutrino events is consistent with being
isotropic across the sky, indicating a strong extragalactic component.
Possible extragalactic sources of cosmic rays and neutrinos are
GRBs
\citep{Waxman1995,Waxman1997,Waxman1999,Dermer2002,Dermer2003,Murase2013},
blazars and non-blazar AGN \citep{Stecker1991,Mannheim1993,
  Mannheim1995,Murase2014}, and starburst galaxies \citep{Loeb2006,Murase2013b}.
Starburst galaxies are also not able to produce PeV neutrino events
\citep{Loeb2006,Tamborra2014}. AGN and GRBs sources are theoretically
able to accelerate protons to sufficiently high energies to produce
$\gtrsim$1\,PeV neutrinos.

\subsection{Gamma-ray Bursts}
Long GRBs can occur with type Ib/c supernovae \citep[for a
  review see][]{Meszaros2006}, though this is not the case for all
long GRBs \citep{vanPutten2009,vanPutten2014}.
The dense iron core of the star undergoing a supernova collapses into a
black hole. Further infalling material forms an accretion disk, which
ejects relativistic jets. These travel outwards rapidly, and interact
with the outer infalling layers. This produces low-energy $\gamma$
rays. As this jet travels further outward it can interact (and be
slowed) by the ambient medium which produces the afterglow at X-ray,
optical and radio wavelengths.
It is unclear whether protons are accelerated in
this relativistic outflow and whether high-energy neutrinos are
produced. It has been shown by \cite{Becker2010} we only expect a low
flux of neutrinos from GRBs, inconsistent with the observed PeV
signal.
LAT results confirm this and suggest that GRBs cannot be the dominant
source of the IceCube neutrinos \citep{Bechtol2015}.
This is also in agreement with an All-Sky search by \cite{Aartsen2016}.
The GRB contribution could be higher for a large number of hypernovae
\citep{Senno2016}. No prompt neutrino event is currently consistent
with a known GRB. Production of neutrinos at PeV-EeV energies
is possible in GRB afterglows.
As there currently is a lack of GRBs that are consistent with the
observed neutrino events, chocked jets and low luminosity GRBs are
being considered \citep{Senno2016,Murase2016}. Chocked jets are
shortened by interaction with the surrounding medium. Weak jets would
produce low luminosities. Either case would be very difficult to
detect in the electromagnetic spectrum and this option cannot be ruled
out currently.

\subsection {Active Galactic nuclei}
Neutrino emission has been predicted for the cores and jets of AGN.
As a correlation between the high-energy EM flux and the neutrino flux is
typically expected, it is likely that bright high-energy emitters
contribute to the flux. Blazars are the dominant class seen by
\Fermi-LAT and the most luminous persistent sources. This makes them
reasonable candidates for contributing to the flux of neutrinos.

Blazars are typically further subdivided into two classes, BL Lacertae
(BL Lacs) and flat-spectrum radio quasars (FSRQs).
Quasars tend to be more luminous than BL Lacs, with their spectra
peaking at lower energies.
Quasars also typically exhibit a strong thermal emission peak in the
optical/UV, called the 'big blue bump' (BBB). This emission feature is
generally interpreted to be the thermal emission from the accretion
disk. And while several problems exist with this classification
\citep[see][]{Lawrence2012}, the accretion disk would provide an adequate photon
seed field for photopion production.
BL Lac objects are often TeV emitter. In addition to the lack of a
photon seed field, this makes it hard to explain the escape of TeV
photons from the dense environments required for photopion production.

Several studies suggest that blazars alone cannot explain all IceCube
contained neutrino events \citep{Feyereisen2016,Gluesenkamp2016}. This is not
incompatible with the PeV events steming from AGN.
\cite{Padovani2014} propose a combination of BL Lac events and PWN for
all of the IC events in three years of data.

We used the six $\gamma$-ray and radio brightest blazars to
estimate the integrated neutrino output. We found that the emission
can be explained by these events, even without including contributions
from fainter blazars or unresolved blazars \citep{Krauss2014}. None of the sources had a
high individual expected neutrino output, which would explain the
events. This work showed that blazars as a class are capable of
producing the observed PeV events.

A follow-up study by ANTARES found two possible neutrino events for
the two sources with the highest expected neutrino output and helped
to constrain the spectral index of the neutrino spectrum \citep{AT2015}.

For the third event at PeV energies, we found one blazar dominating
the expected integrated neutrino output as it was undergoing a
dramatic outburst (a state of high activity with a strongly increased
flux). An increase in em flux is also expected to increase the
neutrino flux. For small time scales this is typically negligible due
to the low chance of detecting neutrinos. This was an outburst that
lasted for a year, and only 8 outbursts reaching a similar fluence
where detected over the lifetime of Fermi, with several of these
before the start of the IceCube experiment.
Unfortunately, we estimate that a 5\% probability of a chance
coincidence remains. It is an interesting result that will be be
confirmed or refuted in the future.
Assuming a physical connection between both, this would constrain the
neutrino velocity to two orders of magnitude better than SN1987A \citep{bb}.

The fourth event was detected at 2.6\,PeV, with a low angular
resolution \citep{IC2016}. Further events are now communicated on
a short time scale via the GCN network and the \Fermi LAT Galactic and
Extragalactic Science Teams coordinated their efforts of searching for
counterparts, but none was found \citep{GCNa,GCNb}. It is worth
noting that the $\gamma$-ray bright blazar PG\,1553+113 is located
close to the position of the neutrino event.

\section{Conclusion}
High-energy neutrino events at $\gtrsim$1\,PeV are promising for
identifying the sources of Ultra-high energy cosmic rays (UHECRs).
Current LAT results suggest that AGN are the most likely candidates,
although GRBs cannot be ruled out completely.
It is reasonable to assume that several source classes contribute to
the observed IceCube events. I find that blazars, as the most
luminous, persistent, and numerous source class at high energies, are
able to calorimetrically explain the observed IceCube events
\citep{Krauss2014}. We further find one blazar in outburst, coincident
with the third PeV event \citep{bb}. This blazar's outburst was one of
the most dramatic outbursts at $\gamma$-ray wavelengths, but also at
radio wavelengths, as observed by TANAMI.
While this association is not statistically significant it is
a tantalizing results. Only better statistics of the neutrino events
will be able to confirm or refute the blazar hypothesis.
It is further necessary to study the non-detection of neutrinos from
other bright blazars is still in agreement with the hypothesis. It is
possible that individual sources have different fractions of
hadronic/leptonic high-energy emission production.

With the construction of the future Km3Net \citep{km3net} array with a
much larger volume and IceCube Gen-2 \citep{gen2a}, a
significant association of a neutrino event and an astrophysical source
will be feasible. It is not unlikely that \Fermi-LAT will still be
operational at this point, providing crucial information about
$\gamma$-ray bright sources. The planned Cherenkov Telescope Array
\citep[CTA;][]{cta}, which will detect $\gamma$-rays at TeV energies, will be useful
in supporting \Fermi-LAT observations and further constraining the spectral shape of
$\gamma$-ray sources.

\begin{acknowledgments}
I thank D. Berge for the helpful comments. I acknowledge funding from
the European Union’s Horizon 2020 research and innovation programme
under grant agreement No. 653477.
The \textit{Fermi}-LAT Collaboration acknowledges support for LAT
development, operation and data analysis from NASA and DOE (United
States), CEA/Irfu and IN2P3/CNRS (France), ASI and INFN (Italy), MEXT,
KEK, and JAXA (Japan), and the K.A.~Wallenberg Foundation, the Swedish
Research Council and the National Space Board (Sweden). Science
analysis support in the operations phase from INAF (Italy) and CNES
(France) is also gratefully acknowledged.
\end{acknowledgments}

\scriptsize
\setlength{\bibsep}{0pt plus 0.3ex}

\nocite{*}
\bibliographystyle{jwaabib}
\bibliography{mnemonic,aa_abbrv,bla}
\end{document}